\documentclass[journal,twocolumn]{IEEEtran}
\usepackage{amsfonts, amsmath, graphicx}
\usepackage{algorithm,algpseudocode,cite}
\usepackage{subfigure}
\usepackage{epstopdf}
\usepackage{ textcomp }
\include{def}

\def\vec{{\mathrm{vec}}}
\def\[{\left[}
\def\]{\right]}
\def\a{\mathbf{a}}
\def\b{\mathbf{b}}
\def\e{\mathbf{e}}
\def\n{\mathbf{n}}
\def\p{\mathbf{p}}
\def\r{\mathbf{r}}
\def\s{\mathbf{s}}
\def\v{\mathbf{v}}
\def\y{\mathbf{y}}

\def\A{\mathbf{A}}
\def\B{\mathbf{B}}
\def\D{\mathbf{D}}
\def\H{\mathbf{H}}
\def\R{\mathbf{R}}
\def\X{\mathbf{X}}

\def\Im{\text{Im}}
\def\Re{\text{Re}}

\def\Gam{\boldsymbol{ \Gamma }}
\def\Sig{\boldsymbol{ \Sigma }}

\newtheorem{lemma}{Lemma}[section]

\newtheorem{remark}{Remark}[section]

\begin{document}
\title{A Block Alternating Optimization Method for Direction-of-Arrival Estimation with Nested Array}
\author{Yunmei Shi, Xing-Peng~Mao, \IEEEmembership{Member,~IEEE,} Chunlei Zhao, and Yong-Tan Liu, \IEEEmembership{Senior Member,~IEEE}
	\thanks{
		This work was supported by the Key Program of National Natural Science Foundation of China (No.61831009).\par		
		Y. Shi, X.-P. Mao, C. Zhao and Y.-T. Liu are with the School of Electronics and Information Engineering, Harbin Institute of Technology, Harbin 150001, China, and the Key Laboratory of Marine Environmental Monitoring and Information Processing, Ministry of Industry and Information Technology, Harbin 150001, China (email: shiyunmeidamei, mxp, zhaochunlei, liuyt@hit.edu.cn).
	}
}
\maketitle

\begin{abstract}
In this paper, direction-of-arrival estimation using nested array is studied in the framework of sparse signal representation.  With the vectorization operator, a new real-valued nonnegative sparse signal recovery model which has a wider virtual array aperture is built. To leverage celebrated compressive sensing algorithms, the continuous parameter space has to be discretized to a number of fixed grid points, which inevitably incurs modeling error caused by off-grid gap. To remedy this issue, a block alternating optimization method is put forth that jointly estimates the sparse signal and refines the locations of grid points. Specifically, inspired by the majorization minimization, the proposed method iteratively minimizes a surrogate function majorizing the given objective function, where only a single block of variables are updated per iteration while the remaining ones are kept fixed. The proposed method features affordable computational complexity, and numerical  tests corroborate its superior performance relative to existing alternatives in both overdetermined and underdetermined scenarios.

%it does not require the user to make any difficult selection of hyperparameters This paper revisits the problem of direction-of-arrival (DOA) estimation in presence of impulsive noise/outliers.  The conventional DOA estimation algorithms which are explicitly or implicitly developed under the Gaussian noise assumption fail to suppress the outliers effectively and thus suffer from severe performance degradation. Inspired by the relaxation technique and the $\ell_p$-fitting criterion,  we propose a BSUM-RELAX method which is robust to the impulsive noise.   BSUM-RELAX decouples the multiple sources optimization problem into a series of $\ell_p$-norm based single source optimization subproblems and performs the parameter estimation alternatively, i.e., it updates the parameters of one source at a time by fixing those of the other sources.  By resorting to the block successive upper-bound minimization (BSUM) approach, a surrogate function which provides a rewighted quadratic upperbound of the $\ell_p$-norm based objective function is devised for two-block parameters, i.e., signal and DOA, estimation. Moreover, we employ the Newton's method to iteratively refine the DOA candidates.  Convergence analysis of the proposed method for single source scenario is also provided.  Finally, numerical results show that BSUM-RELAX achieves a substantial performance improvement over existing methods in many cases of practical interest. 
\end{abstract}


\section{Introduction}

Direction-of-arrival (DOA) estimation of multiple narrow-band sources is one of the most important issues in signal processing, which has diverse applications ranging from synthetic aperture radar \cite{Ender1999}, wireless communications \cite{Guo2017}  to source localization \cite{Miron2015}.  It is well known that many DOA estimation algorithms  are  confined to the overdetermined scenario where the number of sources is less than the number of sensors \cite{MUSIC1986,Shi2015direction,Shi2019Robust,Gallager2008Prin,epuma}. For example, subspace based approaches such as MUSIC \cite{MUSIC1986}  can only resolve up to $M-1$ sources with an $M$-element uniform linear array (ULA). However, the problem of detecting more sources than  sensors emerges in various practical applications \cite{Hoctor1990unify}. To achieve an increase in the degrees of freedom (DOF), many nonuniform array structures which are capable of resolving more sources than the actual number of physical sensors have been developed \cite{Hoctor1990unify,Piya2010nested}. The nested array is one of the most popular nonuniform array structure since it has closed-form expression for the array configuration. Moreover, in \cite{Piya2010nested}, it was shown that the nested array is able to identify up to $\mathcal{O}(M^2)$ sources with only $M$ physical sensors. 

In recent years, numerous DOA estimation methods which can be applied to nested arrays have been developed \cite{Piya2010nested,Han2013improved,Chen2018sparse, Wang2017coarrays, Yang2016efficient}. Among them, the spatial smoothing based MUSIC (SS-MUSIC) \cite{Piya2010nested}  is the most successful subspace method, which achieves significant DOF increase by making the most of the array structure to construct an augmented covariance matrix.  However, SS-MUSIC fails to work well when the number of snapshots is small, or the signal-to-noise ratio (SNR) is low. This is the well known bottleneck shared by all subspace based methods.  Another line of research builds on the emerging technique of compressive sensing (CS), which has benchmarked the performance in diverse signal recovery applications \cite{Wang2018sparse,Hwang2016multivariated,Qaisar2013compressive}, and has also greatly promoted the rise of CS based DOA estimators \cite{Malioutov2005,Stoica2014SPICE, Shen2016Underdetermined, Stoica2015Online}. It was shown that the exploitation of sparsity of the incoming signals helps to improve the performance of the DOA estimators especially in the above mentioned demanding scenarios.  For example, $\ell_1$-SVD \cite{Malioutov2005} was developed via reformulating the measurements in a sparse form with an overcomplete basis consisting of the potential DOA candidates. Then by assuming that all true DOAs exactly lie on the grid points, an $\ell_1$-norm penalty is imposed to locate the DOAs of interest. Though $\ell_1$-SVD exhibits plenty of advantages over the classical methods, which include improved robustness when the SNR is low, the number of snapshot is small, and the correlation of the sources is large, it suffers from the problem of choosing hyper-parameters.  In \cite{Stoica2011sparse}, Stoica \emph{et al.} proposed the sparse iterative covariance-based estimation method (SPICE) by utilizing a sparse covariance fitting criterion, which circumvents the issue of hyper-parameters and yields good resolution performance. 

In practice, both the $\ell_1$-SVD and SPICE have grid mismatches problem since the true DOAs are not always exactly on the sampling grid. On one hand, a dense sampling grid can reduce the gap between the ture DOA and its nearest grid point. On the other hand,  a dense sampling grid increases the computational complexity and the mutual coherent between the columns in the overcomplete basis.
To circumvent the grid mismatch problem, an off-grid model for DOA estimation has been developed \cite{Chen2018sparse, Liu2017,Yang2013,Das2017, Cai2017Efficient}. In \cite{Liu2017},  a new dictionary model based on the first-order Taylor approximation was employed to take the off-grid DOA information into account. It was shown that the new model is able to  achieve higher modeling accuracy. With such dictionary model,  Cai \emph{et al.} proposed a so-called Capon-SPICE (C-SPICE)  which improves the estimation accuracy by estimating the on-grid angles and deviations of the off-grid DOAs independently.  In addition to these efforts, sparse Bayesian learning (SBL) based methods were proposed to iteratively refine the grid-points by viewing the sparse represented signals as hidden variables \cite{Yang2013,Das2017}.  The SBL strategy with nested array was first addressed in \cite{Yang2016efficient} where a linear transformation operation was adopted to eliminate the effect of noise. However, such operation  leads to a reduced working array aperture. To circumvent this problem, Chen \emph{et al.} developed  a new Bayesian inference learning model \cite{Chen2018sparse} which take the noise as a part of the unknown variables to be estimated. To further overcome the grid mismatch issue, Yang \emph{et al.} proposed gridless DOA estimation methods \cite{Yang2014discretization,Yang2016exact} which directly operate in the continuous angle domain and thus avoid the angle discretization problem.  However, so far, these methods can only be applied to the linear array case. 

In this paper, we address the DOA estimation problem from a super-resolution CS perspective, where the nested array is exploited to provide more DOF. Specifically,  by vectorizing the sample covariance of the array data and exploiting its real-valued conditional distribution, we propose an iteratively reweighted method for joint dictionary parameter learning and sparse signal recovery.  The proposed method is developed by employing a block alternating optimization strategy, which judiciously recasts the original problem into a series of successive block minimization subproblems. Different from the original hard instance,  the proposed method updates only a single block of variables every iteration. We show that each subproblem can be tackled using  simple yet effective algorithms. In particular,  to update the sparse signal variance,  we employ the majorization  minimization (MM) technique to construct a surrogate function which upper bounds the original non-convex objective. Rather than solving the surrogate function directly, we use a iterative cyclic minimization technique to further enhance  computational efficiency. As for the update for the noise term, instead of only taking a triming virtual signal vector into account \cite{Yang2016efficient}, we adopt the whole signal vector for DOA estimation and take the noise term as a separate block which can be updated in closed-form. In this way, the whole working array aperture is exploited to provide more accurate DOA estimates. To alleviate the modeling error of the first-order Taylor expansion used in \cite{Chen2018sparse}, we propose a new grid refining procedure by treating the locations of grid points as a block of adjustable variables, for which gradient descent is employed to further refine the potential DOAs. 

The present work builds on but considerably extends \cite{Chen2018sparse} which highlights that the perturbation of the vectorized covariance matrix follows an asymptotic complex Gaussian distribution, and an off-grid SBL method was proposed based on the Gaussian distribution. However, on one hand, its resolution relies on the number of grid points, and the resulting computational complexity is  high when we choose a dense sampling grid. On the other hand, it requires a careful selection of user-defined  hyper-parameters, wherein the  hyper-parameters play a key role of controlling the sparsity of the solution. Unlike the off-grid SBL method  in \cite{Chen2018sparse}, we address the DOA estimation problem in a super-resolution block alternating optimization framework. Specifically, we recast the objective into real-valued formulation, and hence the optimization will be carried out with real-valued operations. To further reduce the computational complexity,  a pruning operation is used to remove these small coefficients together with the associated grid points during the update procedure. Note that the proposed method is able to work without any \emph{a priori} knowledge on the number of sources, and can work well when the number of sources is larger than the number of physical sensors. Moreover, different from the off-grid SBL method which requires the user to select or tune critical hyper-parameters, our method adapts to the signal model via a real-valued conditional distribution automatically. Consequently, it obviates the need for selecting or tuning any hyper-parameter.  Simulation results show that the proposed method yields superior DOA estimation performance than state-of-the-art algorithms.

The rest of the paper is organized as follows. Section II presents the problem formulation. The novel DOA estimation method is developed in Section III. Numerical results are presented in Section IV and conclusions are drawn in Section V. 

\noindent {\bf Notation:} Throughout the paper, we use boldface lowercase letters for vectors and boldface uppercase letters for matrices. Superscripts $(\cdot)^{*}$,  $(\cdot)^T$,  $(\cdot)^H$ and $(\cdot)^{-1}$ represent conjugate, transpose, conjugate transpose and  inverse  respectively. The $\mathbb{E}\{\cdot\}$  stands for mathematical expectation,  $\text{diag}\{\cdot\}$ means forming the given vector as a diagonal matrix and $\vec\{\cdot\}$ is the vectorization operator. The $\odot$ and $\otimes$ are the Khatri-Rao product and Kronecker product, respectively. The $\|\cdot\|_2$ is the $\ell_2$-norm and  $\boldsymbol{I}_{M}$ is the $M\times M$ identity matrix. The $|\cdot|$ represents the absolute value of a scalar or the determinant of a matrix.   The  $\Re(\cdot)$  and  $\Im(\cdot)$ denote  the real part and the imaginary of its argument, respectively. 
%------------------

\section{Problem Formulation}
\begin{figure}
	\centering
	\includegraphics[width=0.85\linewidth]{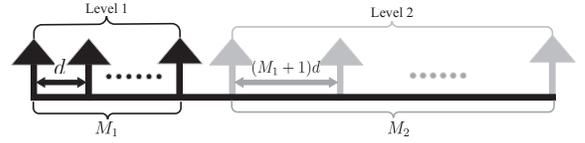}
	\caption{A 2 level nested array with two ULAs.}
	\label{nested_array}
\end{figure}

Consider a nonuniform array with $M$ omnidirectional sensors which are located at $\mathcal{L} = \{0, d_1 \ldots, d_{M-1}\}$, where $d_m$ represents the distance between the $m$-th sensor and the reference one (location 0).  For example, Fig. \ref{nested_array} shows a two level nested array which can be decomposed into two concatenated ULAs, where the inner ULA consists of $M_1$ sensors with inter-element spacing $d$ while the outer ULA has $M_2$ sensors with inter-element spacing $(M_1 + 1)d$. Note that $d$ is usually set as $\lambda/2$ with $\lambda$ being the wavelength.  Mathematically, we take the first sensor of the inner ULA as the reference and get the  location of each sensor as follows
\begin{align}
	\mathcal{L} = &\{ 0, d, \cdots, (M_1 - 1 )d, M_1 d, 2(M_1 + 1)d -d, \cdots,\notag \\
	               & M_2(M_1+1)d-d \}.
\end{align}

Assume that $K$ narrowband sources from directions $\boldsymbol{\theta} = [\theta_1, \theta_2, \cdots, \theta_K]$ impinge on the array.  The array output observations $\y(t) \in \mathbb{C}^{M\times 1}, t=1, 2,\cdots,T$,  can be  modeled as
\begin{align}\label{model1}
\y(t) = \A(\boldsymbol{\theta})\s(t) + \n(t)
\end{align}
where $\A(\boldsymbol{\theta}) = [\a(\theta_1), \a(\theta_2),\cdots, \a(\theta_K)]$, with $\a(\theta_i)$ being the steering vector of the $i$-th source; $T$ is the number of snapshots; and $\n(t) \in \mathbb{C}^{M\times 1}$ is additive noise,  which  is assumed to follow a Gaussian distribution with zero mean and covariance matrix $\sigma_n^2 \boldsymbol{I}_M$. We also assume that the sources are circularly-symmetric Gaussian distributed, which are uncorrelated with each other, i.e., $\R_s = \mathbb{E}\{ \s(t)\s^{H}(t) \} = \text{diag}\{\boldsymbol{\rho}\}$ with $\boldsymbol{\rho} = [\sigma_{s1}^2, \sigma_{s2}^2, \cdots, \sigma_{sK}^2]^{T}$, and are uncorrelated with the noise. Thus the covariance matrix of the observations $\y(t)$ can be written as
\begin{align}
\R_y= \mathbb{E}\{ \y(t)\y^{H}(t) \} = \A(\boldsymbol{\theta})\R_s\A^{H}(\boldsymbol{\theta}) + \sigma_n^2 \boldsymbol{I}_M.
\end{align}
By vectorizing the above covariance matrix $\R_y$, we have
\begin{align}
\r = \vec{(\R_y)} = (\A^{*}(\boldsymbol{\theta})\odot \A(\boldsymbol{\theta})) \boldsymbol{\rho} + \sigma_n^2\boldsymbol{1}_n
\end{align}
where $\A^{*}(\boldsymbol{\theta})\odot \A(\boldsymbol{\theta}) = [\a^{*}(\theta_1)\otimes \a(\theta_1), \cdots, \a^{*}(\theta_K)\otimes \a(\theta_K)]$ and $\boldsymbol{1}_n = [\e_1^{T}, \e_2^{T}, \cdots, \e_M^{T}]^{T}$ with $\e_i$ being a vector with all zeros except for the $i$-th element being $1$. 

In practice, since the covariance matrix $\R_y$ is unavailable, we usually choose to use the sample covariance matrix
\begin{align}\label{hat_R}
\hat{\R}_y = \sum_{t= 1}^{T} \y(t)\y^{H}(t)
\end{align}
instead. It was revealed in \cite{Gallager2008Prin} that when the sources are circularly-symmetric Gaussian distributed, the vectorized sample covariance matrix, i.e., $\hat{\r}= \vec(\hat{\R}_y)$, satisfies the following asymptotic complex Gaussian distribution
\begin{align}
	\hat{\r} \sim \mathcal{CN}(\r, \frac{1}{T} \R^{T}_y\otimes \R_y). 
\end{align} 
We define $\boldsymbol{\epsilon} := \hat{\r} - \r$, then we have 
\begin{align}\label{hat_r}
\hat{\r} = \vec{(\hat{\R}_y)} = (\A^{*}(\boldsymbol{\theta})\odot \A(\boldsymbol{\theta})) \boldsymbol{ \rho} + \sigma_n^2\boldsymbol{1}_n + \boldsymbol{\epsilon}.
\end{align} 
In order to estimate DOAs from \eqref{hat_r}, numerous techniques are proposed in the literature. Among them, one of the most popular techniques that are widely used might be the spatial smoothing (SS) \cite{shan1985spatial}. Combined with the SS technique, different variants of subspace methods, e.g., MUSIC and ESPRIT, are developed. However, these algorithms are able to work at the expense of losing array aperture. To tackle this issue, the CS based technique is tailored in \cite{Yang2016efficient} for nested array. More specifically, the $\hat{\r}$ in \eqref{hat_r} is reformulated in a sparse form using an overcomplete dictionary which is built upon discretized angle set. Assuming $\boldsymbol{\phi} = [\phi_1, \phi_2, \cdots, \phi_N]$ being the grid points of the potential angle set, $\hat{\r}$ can be represented as follows
\begin{align}\label{hat_r_sparse}
\hat{\r}  & = \[ \A^{*}(\boldsymbol{\phi})\odot \A(\boldsymbol{\phi}) \] \p + \sigma_n^2\boldsymbol{1}_n + \boldsymbol{\epsilon} \notag \\
		  & = \B(\boldsymbol{\phi}) \p   + \sigma_n^2\boldsymbol{1}_n + \boldsymbol{\epsilon}
\end{align} 
where $\B(\boldsymbol{\phi}) = [\a^{*}(\phi_1)\otimes \a(\phi_1), \cdots, \a^{*}(\phi_N)\otimes \a(\phi_N)]$ and $\p = [\sigma_{p1}^2, \sigma_{p2}^2, \cdots, \sigma_{pN}^2]^{T}$. To get an overcomplete dictionary $\B(\boldsymbol{\phi}) \in \mathbb{C}^{M^2 \times N}$, we usually have $M^2 \ll N$. Note that in practical applications, the true DOAs do not necessarily lie on the grid points. As a result, the  well studied off-grid DOA estimators are developed to handle this issue \cite{Yang2016efficient,Chen2018sparse}.

\section{Proposed Algorithm}
\subsection{Real-valued Representation of Objective Function}
 As $\p$ is real-valued and nonnegative, it is easy to represent the complex-valued  $\hat{\r}$ in \eqref{hat_r_sparse} in the  following real-valued formulation
\begin{align}\label{hat_r_real}
\bar{\r}  & =  \bar{\B} \p + \bar{\boldsymbol{\epsilon}}.
\end{align}
where $\bar{\r} = [\Re(\hat{\r})^{T} - \sigma_n^2\boldsymbol{1}_n^{T}, \Im(\hat{\r})^{T} ]^{T}$, $\bar{\boldsymbol{\epsilon}} = [\Re(\boldsymbol{\epsilon})^{T}, \Im(\boldsymbol{\epsilon})^{T}]^{T}$,  $\bar{\B} = [\Re(\B(\boldsymbol{\phi}))^{T},  \Im(\B(\boldsymbol{\phi}))^{T}]^{T}$. Since we assume that the incident signals are circularly-symmetric Gaussian distribution,  according to \cite{Gallager2008Prin},  $\bar{\boldsymbol{\epsilon}}$ has the following real-valued Gaussian distribution
\begin{align}
\bar{\boldsymbol{\epsilon}}\sim \mathcal{N}(\boldsymbol{0}, \bar{\R})
\end{align}
with
\begin{equation}
\bar{\R} = \frac{1}{2}
\left[
\begin{array}{cc}
\Re(\frac{1}{T} \R^{T}_y\otimes \R_y) & -\Im(\frac{1}{T} \R^{T}_y\otimes \R_y)\\
\Im(\frac{1}{T} \R^{T}_y\otimes \R_y) & \Re(\frac{1}{T} \R^{T}_y\otimes \R_y)
\end{array}
\right].
\end{equation}
In the following context, we make $\R^{T}_y\otimes \R_y \approx \hat{\R}^{T}_y\otimes\hat{\R}_y$, i.e., it is estimated by the sample covariance matrix.  

Inspired by \cite{Liu2013sparsity}, we assume that $\p \sim \mathcal{N}(\boldsymbol{0}, \boldsymbol{\Gamma})$ with $\boldsymbol{\Gamma} = \text{diag}\{ \boldsymbol{\gamma} \}$ and $\boldsymbol{\gamma} = [\gamma_1, \gamma_2, \cdots, \gamma_{N}]^{T}$. Then the likelihood function of $\bar{\r}$ is given by 
\begin{equation}\label{f_bar_r}
\mathcal{L}(\bar{\r}) = \frac{1}{ |\pi\boldsymbol{ \Sigma_{\bar{\r}}}| }e^{-\bar{\r}^{T}\boldsymbol{ \Sigma_{\bar{\r}} }^{-1} \bar{\r}}
\end{equation}
where $\Sig_{\bar{\r}}  = \bar{\B} \boldsymbol{ \Gamma } \bar{\B}^{T} + \bar{\R} $. Taking the negative logarithm of \eqref{f_bar_r} and neglecting the uninteresting constants, one can obtain the following optimization
\begin{align}\label{L_neg_log}
&\min_{\boldsymbol{ \phi }, \boldsymbol{\gamma}, \sigma_n^2}  \ln| \Sig_{\bar{\r}} | + \bar{\r}^{T} \Sig_{\bar{\r}}^{-1} \bar{\r} \notag \\
&\quad \text{s.t.} \quad \boldsymbol{ \gamma }\geq \boldsymbol{0}, \sigma_n^2 \geq 0.
\end{align}
Intuitively, it is easy to see that we can optimize the objective with respect to ( w.r.t.) the three blocks, i.e., $\boldsymbol{ \phi }, \boldsymbol{\gamma}, \sigma_n^2$, in an alternating optimization fashion. 

\subsection{Algorithm Development}
First of all,  we discuss  the update for $\boldsymbol{\gamma}$ by fixing $\boldsymbol{ \phi }$ and $\sigma_n^2$.  We note that the first term in \eqref{L_neg_log} is a concave function whereas the second term is a convex function of $\boldsymbol{ \gamma}$. It implies that the first term in \eqref{L_neg_log} plays a sparsity-inducing role. We note that the concave function $\ln(x)$ is majorized by its tangent plane at any given point $\hat{x}$. To make the problem in \eqref{L_neg_log} solvable, we apply the MM procedure to linearize  the first term $\ln|\boldsymbol{ \Sigma_{\bar{\r}}}|$. Given $\{ \hat{\gamma}_k \}$, the linear surrogate function of $\ln|\boldsymbol{ \Sigma_{\bar{\r}}}|$ can be written as follows
\begin{align}
f(\boldsymbol{ \gamma }| \hat{\boldsymbol{ \gamma }} ) &= \ln|\hat{\Sig}_{\bar{r}}| + \sum_{k = 1}^{N} \text{tr}\[ \hat{\Sig}_{\bar{r}}^{-1} \bar{\b}_{k}\bar{\b}_{k}^{T} \] (\gamma_k - \hat{\gamma}_k)    \notag \\
& =  \sum_{k = 1}^{N} \bar{\b}_{k}^{T} \hat{\Sig}_{\bar{r}}^{-1} \bar{\b}_{k}\gamma_k + \text{c}  \notag \\
& :=  \sum_{k = 1}^{N} w_k \gamma_k + \text{c}.
\end{align}
with $\text{c}$ representing the uninteresting constants.
Here, we define
\begin{align}\label{eq:w_k}
w_k := \bar{\b}_{k}^{T} \hat{\Sig}_{\bar{r}}^{-1} \bar{\b}_{k} 
\end{align}
where $\bar{\b}_{k}$ is the $k$-th column of $\bar{\B}$,  and $\hat{\Sig}_{\bar{\r}}  = \bar{\B} \hat{\Gam} \bar{\B}^{T} + \bar{\R} $ with $\hat{\Gam} = \text{diag}\{\hat{\gamma}_1, \hat{\gamma}_2, \cdots, \hat{\gamma}_{N} \}$. We can easily verify that 
\begin{align}
f(\boldsymbol{ \gamma }| \hat{\boldsymbol{ \gamma }} ) - \ln| \Sig_{\bar{\r}} | \geq 0
\end{align}
at any given point and the equality holds if and only if $ \boldsymbol{ \gamma } = \hat{\boldsymbol{ \gamma }} $. Consequently, ignoring the terms which are independent of $\boldsymbol{\gamma}$, optimizing  the subproblem w.r.t. $\boldsymbol{ \gamma }$ now reduces to the following problem
\begin{align}\label{L_neg_log2}
&\min_{\boldsymbol{\gamma}} \sum_{k = 1}^{N} w_k \gamma_k  + \bar{\r}^{T} \Sig_{\bar{\r}}^{-1} \bar{\r} \notag \\
&\quad \text{s.t.} \quad \boldsymbol{ \gamma }\geq \boldsymbol{0}.
\end{align}
 Although the objective function in \eqref{L_neg_log2} is convex w.r.t. $\{ \gamma_k \}$, the first-order based methods may overshoot the minimum, and thus not necessarily lead to the monotonic decrease of the  cost. To circumvent this issue, we need the help of the following Lemma \cite{Stoica2014SPICE}:
\begin{lemma}\label{lem:relax_obj}
	Assume $\p = [p_1, p_2, \cdots, p_{N}]^{T}$.
	Then, we have
	\begin{equation}\label{subproblem1}
	\bar{\r}^{H} \Sig_{\bar{\r}}^{-1} \bar{\r}  = \min_{\p}~(\bar{\r} - \bar{\B}\p)^{T}\bar{\R}^{-1}~(\bar{\r} - \bar{\B}\p) + \sum_{k = 1}^{N} |p_k|^2/\gamma_k
	\end{equation}
	and the unique minimizer is
	$$
	\p_{\text{opt}} = \Gam\bar{\B}^{T}\Sig_{\bar{r}}^{-1}\bar{\r}.
	$$
\end{lemma}
Note that the above Lemma is slightly different from the one presented in \cite{Stoica2014SPICE}, where the non-diagonal weighting matrix $\bar{\R}$ was modeled as a diagonal one. To make the paper self-contained, we provide the proof of this Lemma in Appendix A.

Lemma \ref{lem:relax_obj} sheds light on a way to handle \eqref{L_neg_log2} in a cyclic minimization manner. In other words, the cost function in \eqref{L_neg_log2} can also be re-expressed as a joint minimization problem as follows
\begin{align}\label{L_neg_log3}
&\min_{\boldsymbol{\gamma}, \p} \sum_{k = 1}^{N} (w_k \gamma_k + |p_k|^2/\gamma_k)  +  (\bar{\r} - \bar{\B}\p)^{T}\bar{\R}^{-1}~(\bar{\r} - \bar{\B}\p) \notag \\
&\quad \text{s.t.} \quad \boldsymbol{ \gamma }\geq \boldsymbol{0}, \p\geq \boldsymbol{0}.
\end{align}
The structure of problem \eqref{L_neg_log3} is nice since it allows us to optimize the cost w.r.t. the two blocks $ \boldsymbol{\gamma}, \p$ in an alternating minimization fashion. In other words, we can update one block each time while keep the other one fixed. We first consider the subproblem w.r.t. $\p$. Given $\boldsymbol{ \gamma }$, the update for $\p$  can be given by the following optimization
\begin{align}\label{L_neg_logp}
&\min_{ \p} \sum_{k = 1}^{N} |p_k|^2/\gamma_k +  (\bar{\r} - \bar{\B}\p)^{T}\bar{\R}^{-1}~(\bar{\r} - \bar{\B}\p) \notag \\
&\quad \text{s.t.} \quad  \p\geq \boldsymbol{0}.
\end{align}
It follows from Lemma \ref{lem:relax_obj} that  optimal $\{ p_k \}$ of \eqref{L_neg_logp} without the constraints can be readily obtained as 
 \begin{align}\label{p_hat}
 \hat{p}_k = \gamma_k\bar{\b}_{k}^{T}\Sig_{\bar{r}}^{-1}\bar{\r}, k = 1, 2, \cdots, N.
 \end{align} 
 As for the constraints $\p\geq \boldsymbol{0}$, it is intuitively obvious that we can project $\p$ onto the nonnegative cone $\mathcal{C} = \mathbb{R}_{+}^{N}$ as follows
 \begin{align}\label{eq:p_k}
 \p = \Pi_{\mathcal{C}}(\hat{\p}) = (\hat{\p})_{+}
 \end{align}
 where the nonnegative part operator $()_{+}$ is taken elementwise. Thus,
 to project onto $\mathcal{C}$, we simply replace each negative component of $\p$
 with zero. 
Similar to  the optimization procedure of $\p$, fixing $\p$, minimization of \eqref{L_neg_log3} w.r.t. $\{ \gamma_k\}$ yields
\begin{align} \label{eq:gamma_k}
\gamma_k = \frac{p_k}{w_k^{1/2}}, k = 1, 2,\cdots, N.
\end{align}
It should be noted that the update for $\p$ and $\boldsymbol{ \gamma}$ is an inner-loop cyclic minimization which targets at locating the optimal $\boldsymbol{ \gamma }$ under the condition that the other block of variables, i.e., $\sigma_n^2$ and $\boldsymbol{ \phi }$, are fixed.

Next, we consider the update for $\sigma_n^2$. To tackle it, the terms which are independent of $\sigma_n^2$ in \eqref{L_neg_log} are ignored, and then this problem turns to the following optimization
\begin{align}\label{L_neg_log_sigma_n}
&\min_{\sigma_n^2}  \bar{\r}^{H} \Sig_{\bar{\r}}^{-1} \bar{\r} \notag \\
&\quad \text{s.t.} \quad \sigma_n^2 \geq 0.
\end{align}
Together with Lemma \ref{lem:relax_obj}, the problem in \eqref{L_neg_log_sigma_n} can be rewritten as 
\begin{align}\label{L_neg_log_sigma_n2}
 &\min_{\sigma_n^2, \p}~(\bar{\r} - \bar{\B}\p)^{T}\bar{\R}^{-1}~(\bar{\r} - \bar{\B}\p) + \sum_{k = 1}^{N} |p_k|^2/\gamma_k \notag \\
 &\quad \text{s.t.} \quad \sigma_n^2 \geq 0, ~\p \geq 0.
\end{align}
Conditioned on the estimates of $\p$ given in \eqref{eq:p_k} and after some straightforward calculations, we arrive at the following cost function for updating $\sigma_n^2$
\begin{align}\label{L_neg_log5}
\min_{\sigma_n^2} ~&(\v_1 - \sigma_n^2\boldsymbol{1}_n  )^{T} \Re( \hat{\R}^{-T}_y\otimes \hat{\R}_y^{-1}) (\v_1 - \sigma_n^2\boldsymbol{1}_n ) \notag \\
				 & - (\v_1 - \sigma_n^2\boldsymbol{1}_n)^{T}\Im( \hat{\R}^{-T}_y\otimes \hat{\R}_y^{-1})\v_2 \notag \\
				 \text{s.t.} ~& \sigma_n^2 \geq 0 
\end{align}
where $\v_1 = \Re(\hat{\r})  - \Re(\B)\p$ and $\v_2 = \Im(\hat{\r})- \Im(\B)\p$. The solution to the objective in \eqref{L_neg_log5} is given by  
\begin{align}\label{hat_sigma_n2}
\hat{\sigma}_n^2 = \frac{\boldsymbol{1}_n^{T}(\Re( \hat{\R}^{-T}_y\otimes \hat{\R}_y^{-1})\v_1 - \Im( \hat{\R}^{-T}_y\otimes \hat{\R}_y^{-1})\v_2)}{\boldsymbol{1}_n^{T} \Re( \hat{\R}^{-T}_y\otimes \hat{\R}_y^{-1}) \boldsymbol{1}_n}.
\end{align}
We further take the constraint, i.e., $\sigma_n^2 \geq 0 $, into consideration. It is easy to see that the update for  $\sigma_n^2$ in \eqref{L_neg_log5} can be readily obtained as
\begin{equation}\label{sigma_n2}
\sigma_n^2 =  \left\{
\begin{aligned}
&\hat{\sigma}_n^2,  &\hat{\sigma}_n^2 >0  \\
& \text{unchanged},  &\text{otherwise}
\end{aligned}
\right..
\end{equation}

We now turn to discuss the update procedure for $\boldsymbol{\phi}$. Since the objective function in \eqref{L_neg_log} is highly nonlinear and  non-convex w.r.t. $\boldsymbol{\phi}$, rather than optimizing the objective in \eqref{L_neg_log} directly, we consider using the more simple surrogate function given in \eqref{L_neg_log3} as its cost function. Fixing the other blocks of variables, i.e., $\boldsymbol{ \gamma }, \p$ and $\sigma_n^2$, we deal with the 
following optimization
\begin{align}\label{L_neg_log6}
\min_{\boldsymbol{ \phi }} \sum_{k = 1}^{N} (w_k \gamma_k + |p_k|^2/\gamma_k)  +  (\bar{\r} - \bar{\B}\p)^{T}\bar{\R}^{-1}~(\bar{\r} - \bar{\B}\p).
\end{align}
Substituting \eqref{p:optimal} into the objective in \eqref{L_neg_log6} and ignoring terms independent of $\boldsymbol{\phi}$,
we arrive at
\begin{align}\label{L_neg_log4}
\min_{\boldsymbol{ \phi }} -\bar{\r}^{T} \bar{\R}^{-1} \bar{\B} (\bar{\B}^{T} \bar{\R}^{-1} \bar{\B} + \Gam^{-1})^{-1} \bar{\B}^{T}\bar{\R}^{-1}\bar{\r}
\end{align}
where $\boldsymbol{\phi}$ is nonlinearly embedded in the augmented dictionary term $\bar{\B}$. An analytical solution of the above minimization \eqref{L_neg_log4} is difficult to obtain, because the objective function is non-convex and inherently nonlinear w.r.t. $\boldsymbol{\phi}$. To tackle it, instead of minimizing the objective in \eqref{L_neg_log4} at each iteration, we seek to search for a new estimate such that the cost function is guaranteed to be non-increasing  throughout the whole process. We define
\begin{align}
f(\boldsymbol{\phi}) & :=   -\bar{\r}^{T} \bar{\R}^{-1} \bar{\B} (\bar{\B}^{T} \bar{\R}^{-1} \bar{\B} + \Gam^{-1})^{-1} \bar{\B}^{T}\bar{\R}^{-1}\bar{\r}\notag \\
& :=  -\bar{\r}^{T} \bar{\R}^{-1} \bar{\B} \H \bar{\B}^{T}\bar{\R}^{-1}\bar{\r}
\end{align}
where $\H := (\bar{\B}^{T} \bar{\R}^{-1} \bar{\B} + \Gam^{-1})^{-1} $. Our goal is to search for a new estimate $\boldsymbol{\phi}^{(r+1)}$ such that the following inequality holds
\begin{align}\label{ineq_phi}
	f(\boldsymbol{\phi}^{(r+1)}) \leq f(\boldsymbol{\phi}^{(r)})
\end{align}
where $r$ indicates the $r$-th inner iteration for updating $\boldsymbol{\phi}$. Since our target function $f(\boldsymbol{\phi})$ is differentiable w.r.t. $\boldsymbol{\phi}$,  this motivates us to employ gradient descent method to obtain such an estimate.  Specifically, the update of $\boldsymbol{\phi}$ is given by 
\begin{align}\label{update_DOA}
\boldsymbol{\phi}^{(r+1)} = \boldsymbol{\phi}^{(r)} - \mu^{(r)}\frac{\partial f(\boldsymbol{\phi})}{\partial \boldsymbol{\phi}}|_{\boldsymbol{\phi} = \boldsymbol{\phi}^{(r)}}
\end{align}
where $\mu^{(r)}$ is the  step-size, and the first derivative of  $f(\boldsymbol{\phi})$ w.r.t. $\phi_k$ is as follows
\begin{align}\label{f_der_DOA}
\frac{\partial f(\boldsymbol{\phi})}{\partial \phi_k} = &-\bar{\r}^{T} \bar{\R}^{-1}\big[ - \bar{\B}\H (\D_k^{T}\bar{\R}^{-1} \bar{\B} + \bar{\B}^{T}\bar{\R}^{-1}\D_k)\H\bar{\B}^{T}\notag \\
 &+ \D_k \H \bar{\B}^{T} + \bar{\B}\H\D_k^{T} \big] \bar{\R}^{-1} \bar{\r}
\end{align}
with $\D_k = \frac{\partial \bar{\B}}{\partial\phi_k}$. Details of the derivation of \eqref{f_der_DOA} are provided in Appendix B.

Due to the fact that the $\p$ in \eqref{L_neg_log6} is a sparse signal, it is not necessary to fully refine the set $\boldsymbol{\phi}$. Instead, it is only required to further refine these elements of $\boldsymbol{\phi}$ whose corresponding elements in $\p$ are nonzeros.  Therefore, by pruning the current grid points set before optimizing over it, the computational complexity is decreased with the reduction in the dimension of active grid points considered. Specifically, we prune the current grid points by thresholding the elements in $\p$. It can be seen from \eqref{L_neg_log6} that when $p_k$ is small enough, the contributions of the corresponding $k$-th grid point of the current dictionary $\bar{\B}$ in synthesizing the original signal is negligible, and thus it is justified to remove this point from the current grid point set. Mathematically, we proceed to prune the grid point set through 
\begin{align}\label{phi_Omega}
\boldsymbol{\phi}_{\Omega} = \boldsymbol{\phi}(\{ k| p_k\geq \delta_{th} \})
\end{align}
where  $\Omega \subset \{ 1, 2, \cdots, N\}$ denotes the set of indices of the pruned grid points and $\delta_{th}$ is the threshold used to measure the size of $p_k$. Note that along with the pruning of grid points, the associated two blocks, i.e., $ \boldsymbol{\gamma}$ and $ \p$, should be pruned accordingly as well. 

\subsection{Outline of the Proposed Method and some Complementary Issues}
For clarification, we summarize the procedures of the proposed method in \textbf{Algorithm} \ref{algorightm1}.

%------------------------------------
\begin{algorithm}[ht]
	\caption{Proposed Algorithm.}
	\begin{algorithmic}[1]
		\State \textbf{Input:} $\{ \y(t) \}_{t=1}^{T}$, $ \boldsymbol{ \phi }_0 $ and $\boldsymbol{ \gamma }^0$ (see \emph{Remark ~\ref{remark1}}). 
		\State \textbf{Initialization:} $\boldsymbol{\phi}_{\Omega} = \boldsymbol{\phi }_0$,  $\boldsymbol{ \gamma } = \boldsymbol{ \gamma }^{0}$.
		\While{stopping criterion has not been reached}
		\State Update the elements of $\p$  via \eqref{p_hat} and \eqref{eq:p_k};
		\State Update the elements of  $\boldsymbol{ \gamma }$ via \eqref{eq:w_k} and \eqref{eq:gamma_k};
		\State Update $\sigma_n^2$ via \eqref{hat_sigma_n2} and \eqref{sigma_n2};
		\State Prune the current grid points set to $\boldsymbol{\phi}_{\Omega}$ via \eqref{phi_Omega};
		\State Prune the sparse vector $\p$ and $\boldsymbol{\gamma}$ accordingly;
		\State Update the current grid points via \eqref{update_DOA} and \eqref{f_der_DOA}; 
		\State Update the dictionary $\bar{\B}$ using the refined grid point set. Also replace  $\p$ and $\boldsymbol{ \gamma }$ with the pruned ones, respectively;			
		\EndWhile
		\State \textbf{Output:} $\boldsymbol{\phi}_{\Omega}$ and $\p$.
	\end{algorithmic}\label{algorightm1}
\end{algorithm}

\begin{remark}\label{remark1}
	We can initialize $\boldsymbol{\phi}$ with a uniform sampling grid as follows
	\begin{align}
	\boldsymbol{\phi}_{0} = \frac{180^{\circ}}{N}[0, 1, \cdots, N-1].
	\end{align}
	As for the initialization of $\boldsymbol{ \gamma }$, it is well known that there are many ways to carry out spectral estimation, which can be used to initialize $\boldsymbol{ \gamma }$. However, we should take the computational cost of initialization into consideration. To balance the computational efficiency and estimation accuracy, we choose to initialize $\boldsymbol{ \gamma }$ with the spectrum obtained by the periodogram method \cite{Stoica2014SPICE}. Mathematically, it can be expressed as 
	\begin{align}
		\gamma^{0}_{k} = \frac{|\bar{\b}_k^{T}\bar{\r}|^2}{\| \bar{\b}_{k}\|_2^4}, k = 1, 2, \cdots, N.
	\end{align}
	
\end{remark}

\begin{remark}
	As mentioned earlier, our goal is to search for a new estimate of $\boldsymbol{\phi}$	to meet the condition \eqref{ineq_phi}. To further improve the  reconstruction accuracy, the elements $\{ \phi_k \} $ can be refined in a  sequential manner. Moreover,  we can employ the backtracking line search technique to determine the step-size $\mu^{(r)}$ in \eqref{update_DOA}.  According to our experience, a new estimate which satisfies \eqref{ineq_phi} can be easily obtained in a few iterations. 
\end{remark}

\begin{remark}
	The computational cost of each outer iteration of the proposed method is dominated by the matrix inverse operation, i.e., $\Sig_{\bar{r}}^{-1}$ and $\H$ in updating $\boldsymbol{ \gamma }$ and $\boldsymbol{ \phi }$, respectively. Specifically, the evaluations of $\Sig_{\bar{r}}^{-1}$ and $\H$ cost $\mathcal{O}(M_r^3)$ ($M_r$ is the length of $\bar{\r}$) and $\mathcal{O}(\tilde{N}^3)$ flops, respectively. Note that  $\tilde{N}$ is the number of active grid points at the current iteration and a large $\tilde{N}$ will incur a high computational cost. To circumvent this issue,   when $\tilde{N} > M_r$, we can use the matrix inversion lemma to reduce the computational complexity to $\mathcal{O}(M_r^3)$ \cite{Wipf2004sparse}. Thus the computational complexity of each outer iteration becomes $\mathcal{O}(\min\{ M_r^3, \tilde{N}^3\})$.
\end{remark}

%------------------------------------
%For simple reference, we will use super-resolution block alternating optimization algorithm (SR-BAO) to refer to the  proposed method.
%for OGSBI, because it is slow in the case of a dense sampling grid, we use a coarser grid with $N = 180$ as recommended in \cite{Yang2013}.
% we consider a nested array with $6$ elements to perform DOA estimation. 
\section{Simulation Results}
In this section, we consider a nested array with $6$ elements to perform DOA estimation.  In particular, a $2$-level nested array of $M_1 = M_2 = 3$ sensors with locations  $\mathcal{L} = \{ 0, d, 2d, 3d, 7d, 11d \}$ is used, where $d = \lambda/2$ is the inter-sensor spacing of the inner ULA.
Numerical simulations have been carried out to evaluate the performance of the proposed method. Three other state-of-the-art algorithms, i.e., C-SPICE \cite{Cai2017Efficient}, R-SBL \cite{Chen2018sparse} and SPA \cite{Yang2014discretization}, are also included for comparison.   We employ the CRB given in \cite{Wang2017coarrays} which can be applied to the underdetermined case to provide a benchmark for evaluating the performances of these algorithms. It is  assumed that all the signals have identical powers and the SNR is defined as 
\begin{equation}
\text{SNR} = \frac{\mathbb{E}\{ \|\s(t)\|_2^2 \}}{\sigma_n^2}.
\end{equation}
Two statistical performance measures, i.e.,  root mean square error (RMSE) and probability of resolution (PR), are used. The RMSE is defined as 
\begin{equation}
\text{RMSE} = \sqrt{\frac{1}{200} \sum_{i = 1}^{200} \sum_{k =1 }^{K}( \hat{\theta}_{k, i} - \theta_{k} )^2}
\end{equation}
where it is assumed that $200$ Monte-Carlo tests are performed and $\hat{\theta}_{k, i}$ is the DOA estimates of the $k$-th signal in the $i$-th Monte-Carlo test. The PR is calculated based on the ability to detect $K$ sources within $\delta_{\theta}$ degrees from the true DOAs. In other words,  we call it successes in resolving sources if all the DOA estimates satisfy $|\hat{\theta}_k - \theta_{k}| \leq \delta_{\theta}, k = 1, 2, \cdots, K $, and PR is computed as the ratio of the number of successes and the total number of independent runs. Moreover, all experiments are performed using MATLAB2018a on a system with 3.3GHz  Intel core i3-3220 CPU and 6GB of RAM. 

\subsection{Overdetermined DOA estimation}
\begin{figure}
	\centering
	\includegraphics[width=0.85\linewidth]{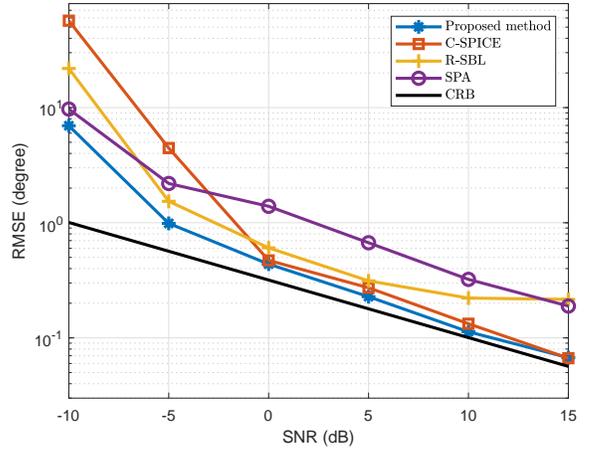}
	\caption{RMSE performances versus SNR for three sources in the overdetermined case.}
	\label{MSE_SNR_S3}
\end{figure}

We first consider the overdetermined scenario where there are three equal-power uncorrelated narrowband sources impinging onto the nested array and their corresponding DOAs are randomly  generated in the range $[-90^{\circ} ~ 90^{\circ}]$.  For the off-grid sparsity-inducing methods, i.e., C-SPICE and R-SBL,  a grid interval of $1^{\circ}$ is used. The number of snapshots is fixed at  $T = 200$. We set $N = 200$ and $\delta_{th} = 0.05$ for the proposed method. Moreover, the stopping criterion is set as either  $\|\p^{\ell+1}-\p^{\ell}\|_2 \leq \eta$ ($\ell$ is the number of outer iterations) or $\ell$ reaching  $\ell_{\text{max}}$, with $\eta = 10^{-6}$ and $\ell_{\text{max}} = 160$ in this scenario. In Fig. \ref{MSE_SNR_S3}, we depict the RMSE of respective algorithm versus SNR which varies from $-10$ dB to $15$ dB.  It is observed from Fig. \ref{MSE_SNR_S3} that our method yields the best performance among all the competitors and outperforms the other algorithms by a big margin especially when SNR $\leq -5$ dB. We see that C-SPICE works well and is only slightly inferior to our method in high SNR scenarios, e.g.,  SNR $\geq 0$ dB.  However, it fails to work efficiently when SNR $\leq 0$ dB.  On the contrary,  R-SBL performs slightly better than C-SPICE when  SNR is low while it cannot converge to the CRB when  SNR $> 10$ dB.  That's potentially because the performance of R-SBL is sensitive to hyper-parameters. 
%Moreover, we find that  ANM has the worst performance . 

\begin{figure}
	\centering
	\includegraphics[width=0.85\linewidth]{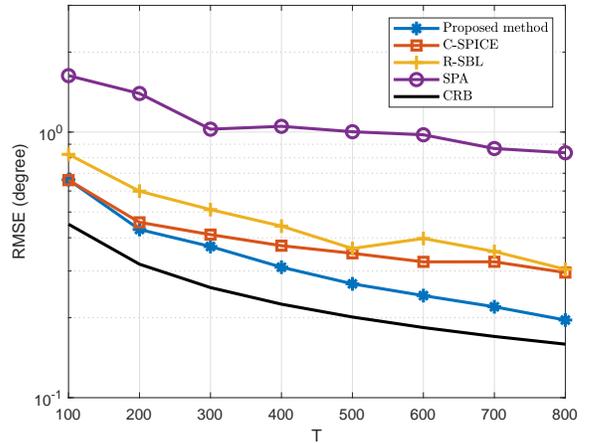}
	\caption{RMSE performances versus $T$ for  three sources in the overdetermined case.}
	\label{MSE_snap_S3}
\end{figure}

Fig. \ref{MSE_snap_S3} shows the RMSEs of the tested algorithms versus the number of snapshots with  SNR $= 0$ dB.  We generate the signals and the grid points in the same way as Fig. \ref{MSE_SNR_S3}.  It is observed that our method achieves the best performance throughout the whole region.  Specifically, the RMSE of our method is approximately equal to that of C-SPICE when $T = 100$. However, the C-SPICE is gradually outperformed by our method as $T$ increases. Note that both our method and the C-SPICE have better performance than the other competitors, i.e., R-SBL and SPA.  We should point out that the SPA does not show very good performance all through, which is possibly caused by the frequency splitting phenomenon \cite{Yang2015gridless}.  

In Table \ref{tab:1}, we investigate the average computation time by varying the number of snapshots $T$. The parameter settings are the same as Fig. \ref{MSE_snap_S3}. We see that the computation times of the four methods fluctuate a little bit as the number of snapshots increases from $100$ to $800$. It means that convergence rate of these methods are insensitive to the number of snapshots. Moreover, it is observed that 
the proposed method is substantially faster than the other competitors throughout the whole region, which indicates that our method  gains significant predominance over the state-of-the-art algorithms in terms of computational complexity.
\begin{table}	
	\caption{Average computation time of the competitors  for  three sources in the overdetermined case (seconds).}
	\label{tab:1}       % Give a unique label
	\begin{tabular}{lcccc}		
		%\toprule
		\hline\noalign{\smallskip}		
		 & Proposed method & C-SPICE & R-SBL & SPA  \\		
		\noalign{\smallskip}\hline\noalign{\smallskip}		
		$T = 100$ &  0.3950 &  0.6591  & 1.8884 &  0.9364  \\		
		$T = 200$ &  0.3493 &  0.7075  & 1.8308 &  1.2643 \\	
		$T = 300$ &  0.4077 &  0.6919  & 1.7066 &  0.9597 \\
		$T = 400$ &  0.5646 &  0.7151  & 1.7339 &  0.9549 \\
		$T = 500$ &  0.3862 &  0.6981  & 1.6839 &  1.0083 \\
	    $T = 600$ &  0.4022 &  0.7859  & 1.8713 &  1.1087 \\
		$T = 700$ &  0.3871 &  0.8005  & 1.8939 &  0.9803 \\	
		$T = 800$ &  0.3183 &  0.6844  & 1.6023 &  0.9423 \\	
		\noalign{\smallskip}\hline
		%\bottomrule		
	\end{tabular}	
\end{table}

\subsection{Underdetermined DOA estimation}
In what follows, we study the performance of the proposed method in  the scenario where there exist more sources than physical sensors. To this end, assume that there are $K =7$ equal-power signals with DOAs $[-54.8^{\circ}, -38.2^{\circ}, -28.6^{\circ}, 3.3^{\circ}, 20.5^{\circ}, 30.6^{\circ}, 48.5^{\circ}]$ impinging onto the 6-element nested array. For our method, a general guideline for choosing $N$ is to let $N \gg K$ such that a finer initial grid can be obtained. As a result, compared with the overdetermined scenario, a denser sampling grid $\frac{180^{\circ}}{N}[0, 1, \cdots, N-1]$ with $N = 300$ is employed to provide finer initial guess for $\boldsymbol{\phi}$. Moreover, a stricter  stopping criterion, i.e., $\eta = 10^{-7}$ and $\ell_{\text{amx}} = 500$,  is used in this case. As for C-SPICE, we increase the number of grid points to $500$,   while for R-SBL, because it is slow in the case of a dense sampling grid, we keep the number of grid points unchanged. 

\begin{figure}
	\centering
	\includegraphics[width=0.85\linewidth]{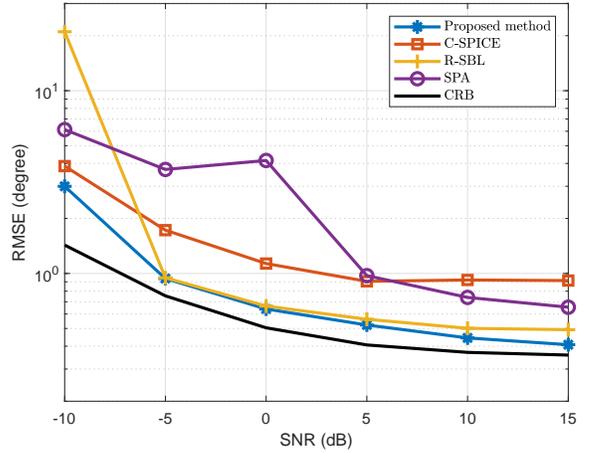}
	\caption{RMSE performances versus SNR for seven sources in the underdetermined case.}
	\label{MSE_SNR_S7}
\end{figure}

Fig. \ref{MSE_SNR_S7} depicts the RMSE performances of the tested algorithms, where the SNR is increased from $-10$ dB to $15$ dB and the number of snapshots is fixed at $500$. We see that the proposed method performs slightly better than  R-SBL  and much better than the other methods, i.e., C-SPICE and SPA, when SNR $\geq -5$ dB.  In the low SNR scenarios, e.g., SNR = $-10$ dB, R-SBL fails to work efficiently while our method has the smallest RMSE and  significantly outperforms the other algorithms. Moreover, it is empirically found that the SBL based algorithms are sensitive to hyper-parameters, which might be the main reason for the bad performance of R-SBL in the low SNR scenarios.  We emphasize that our method is hyper-parameter free and can easily handle the off-grid DOA estimation issue.  
%Note that the RMSE of SPA increases a little bit at the points of $-5$ dB to $0$ dB, which is possibly caused by the frequency splitting phenomenon.

\begin{figure}
	\centering
	\includegraphics[width=0.85\linewidth]{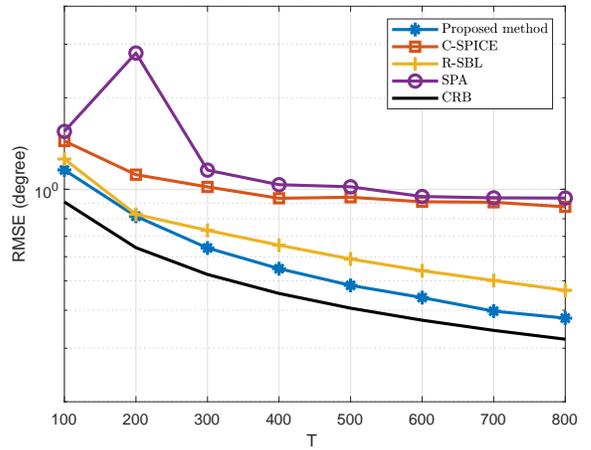}
	\caption{RMSE performances versus $T$ for  seven sources in the underdetermined case.}
	\label{MSE_snap_S7}
\end{figure}

In Fig. \ref{MSE_snap_S7}, we investigate the RMSE performance versus the number of snapshots $T$. We choose to set SNR $ = 5$ dB and increase $T$ from $100$ to $800$. The other parameters keep the same as those in Fig. \ref{MSE_SNR_S7}. It can be seen that the performance of respective algorithm improves along with increasing $T$. In particular, the proposed method and R-SBL outperform the other competitors by a big margin through the whole snapshots  region. Moreover, the RMSE of our method is approximately the same as that of R-SBL at the points of $T= 100$ to $T = 200$ and becomes the smallest one among all the competitors when $T$ increase from $300$ to $800$. We note that the SPA fails to perform well and its RMSE increases a little bit at the points of $100$  to $200$, possibly because it suffers from a more severe frequency splitting problem in this scenario. 

In Table \ref{tab:2}, we tabulate  the computation complexity of the four algorithms  with the increase of $T$. The parameter settings are kept the same as Fig. \ref{MSE_snap_S7}. It is seen that SPA is the fastest one while C-SPICE has the highest computational burden. However, as can be observed from  Fig. \ref{MSE_snap_S7}, both of them fail to work well even when $T$ is large. Unlike the overdetermined scenario, computational cost of the proposed method is higher than that of R-SBL at certain points. This is mainly caused by the fact that our method requires more iterations to meet the stopping criterion. 
%possibly because the number of iterations becomes larger when the number of sources is larger than the number of sensors.

\begin{table}	
	\caption{Average computation time of the competitors for  seven sources in the underdetermined case (seconds).}
	\label{tab:2}       % Give a unique label
	\begin{tabular}{lcccc}		
		%\toprule
		\hline\noalign{\smallskip}		
		& Proposed method & C-SPICE & R-SBL & SPA  \\		
		\noalign{\smallskip}\hline\noalign{\smallskip}			
		$T = 100$ &  2.5225 &  6.7443  & 1.8553 &  1.2914 \\	
		$T = 200$ &  2.3563 &  6.3272  & 1.7231 &  1.1963 \\
		$T = 300$ &  2.1211 &  6.1061  & 1.7215 &  1.2473 \\
		$T = 400$ &  1.8938 &  6.7930  & 1.9309 &  1.3732 \\
		$T = 500$ &  1.5140 &  5.4473  & 1.6605 &  1.1965 \\
		$T = 600$ &  1.9930 &  5.5004  & 1.7678 &  1.2258 \\	
		$T = 700$ &  2.2536 &  4.9373  & 1.7319 &  1.2004 \\
		$T = 800$ &  2.3520 &  4.9479  & 1.6918 &  1.1991 \\	
		\noalign{\smallskip}\hline
		%\bottomrule		
	\end{tabular}	
\end{table}

\begin{figure}
	\centering
	\includegraphics[width=0.85\linewidth]{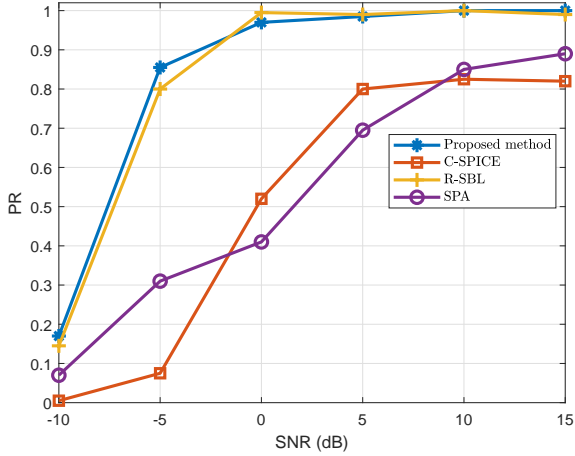}
	\caption{PR performances versus SNR for  seven sources in the underdetermined case.}
	\label{PR_SNR_S7}
\end{figure}

In Fig. \ref{PR_SNR_S7}, we test the impact of SNR on the PR performances of the four algorithms, where $\delta_{\theta}$ is fixed at $0.8^{\circ}$ and the SNR is increased from $-10$ dB to $15$ dB.  The parameter settings are kept the same as Fig. \ref{MSE_SNR_S7}. We see that the PR performance of respective algorithm improves with the increase of SNR. Specifically, the PR performances of our method and R-SBL approximately approach $100\%$ when SNR $\geq 5$ dB, which is much larger than those of C-SPICE and SPA. In the low SNR scenarios, e.g., SNR $<0$ dB,  R-SBL is slightly inferior to our method, while it achieves a little bit larger PR at the point of SNR $=0$ dB. 

\begin{figure}
	\centering
	\includegraphics[width=0.85\linewidth]{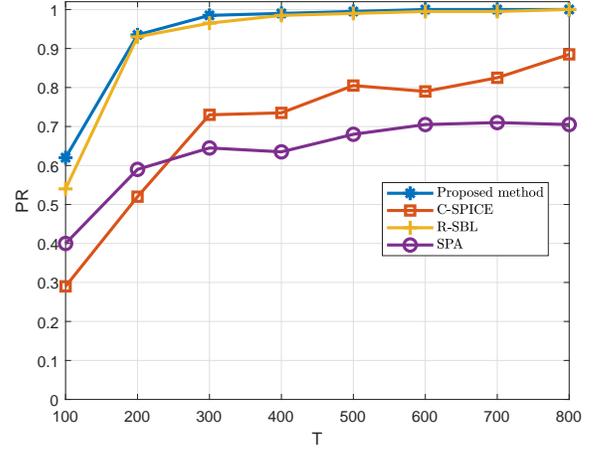}
	\caption{PR performances versus $T$ for  seven sources in the underdetermined case.}
	\label{PR_snap_S7}
\end{figure}

Fig. \ref{PR_snap_S7} depicts the PR performances versus the number of snapshots, where we fix  SNR at $5$ dB and increase the number of snapshots from $100$ to $800$. The remaining parameters are the same as those in Fig. \ref{PR_SNR_S7}. We notice that the proposed method can provide similar performance as R-SBL when $T > 200$, and  outperforms the other two algorithms, i.e., C-SPICE and SPA, by a considerable margin. Moreover, it is seen that our method yields the best performance when the number of snapshots is small, e.g., $T= 100$.

\section{Conclusion}
In this paper, we revisited the DOA estimation problem using nested array, where this problem was studied in a CS framework and the dictionary is characterized by a set of unknown parameters in a continuous domain. By resorting to the MM technique, the original objective was recast into a series of successive block minimization subproblems, resulting in an iterative block alternating optimization algorithm. Numerical results showed that the proposed method can provide reliable DOA estimates and outperform the state-of-the-art algorithms. 

\appendices
\section{Proof of Lemma \ref{lem:relax_obj}}
We rewrite the problem in \eqref{subproblem1} into matrix form as follows
\begin{align}
\min_{\p}~(\bar{\r} - \bar{\B}\p)^{T}\bar{\R}^{-1}~(\bar{\r} - \bar{\B}\p) + \p^{T}\Gam^{-1}\p. 
\end{align}
A simple calculation yields the minimizer
\begin{align}\label{p:optimal}
\p^{*} = (\Gam^{-1} + \bar{\B}^{T}\bar{\R}^{-1} \bar{\B})^{-1}\bar{\B}^{T}\bar{\R}^{-1}\bar{\r}.
\end{align}
Next, by exploiting the fact that 
\begin{align}
	(\Gam^{-1} + \bar{\B}^{T}\bar{\R}^{-1} \bar{\B})\Gam\bar{\B}^{T} 
	& = \bar{\B}^{T} + \bar{\B}^{T}\bar{\R}^{-1} \bar{\B}\Gam\bar{\B}^{T} \notag \\
	& = \bar{\B}^{T}\bar{\R}^{-1}(\bar{\R} + \bar{\B}\Gam\bar{\B}^{T})\notag \\
	& = \bar{\B}^{T}\bar{\R}^{-1}\Sig_{\bar{\r}}
\end{align}
we have 
\begin{align}\label{Gam:equality}
\Gam\bar{\B}^{T}\Sig_{\bar{\r}}^{-1} = (\Gam^{-1} + \bar{\B}^{T}\bar{\R}^{-1} \bar{\B})^{-1}\bar{\B}^{T}\bar{\R}^{-1}.
\end{align}
Substituting \eqref{Gam:equality} into \eqref{p:optimal} yields
\begin{align}\label{p:optimal2}
\p^{*} = \Gam\bar{\B}^{T}\Sig_{\bar{\r}}^{-1}\bar{\r}.
\end{align}
Finally, we evaluate the objective at $\p^{*}$. Since 
\begin{align}
\bar{\r} - \bar{\B}\p^{*} &= \bar{\r} - \bar{\B}\Gam\bar{\B}^{T}\Sig_{\bar{\r}}^{-1}\bar{\r} \notag \\
& = (\Sig_{\bar{\r}} - \bar{\B}\Gam\bar{\B}^{T})\Sig_{\bar{\r}}^{-1}\bar{\r} \notag \\
& = \bar{\R}\Sig_{\bar{\r}}^{-1}\bar{\r}
\end{align}
and then we arrive at 
\begin{align}
&(\bar{\r} - \bar{\B}\p^{*})^{T}\bar{\R}^{-1}~(\bar{\r} - \bar{\B}\p^{*}) + \p^{*T}\Gam^{-1}\p^{*}\notag \\
& = \bar{\r}^{T}\Sig_{\bar{\r}}^{-1}\bar{\R} \Sig_{\bar{\r}}^{-1}\bar{\r} + \bar{\r}^{T}\Sig_{\bar{\r}}^{-1}\bar{\B} \Gam \bar{\B}^{T} \Sig_{\bar{\r}}^{-1}\bar{\r} \notag \\
& = \bar{\r}^{T}\Sig_{\bar{\r}}^{-1} (\bar{\R} + \bar{\B} \Gam \bar{\B}^{T})\Sig_{\bar{\r}}^{-1}\bar{\r}\notag \\
& = \bar{\r}^{T}\Sig_{\bar{\r}}\bar{\r}
\end{align}
which completes the proof.

\section{Derivative of $f(\boldsymbol{\phi})$  w.r.t. $\boldsymbol{\phi}$}
We define 
\begin{align}
\X := \bar{\B} \H \bar{\B}^{T}
\end{align}
with
\begin{align}
\H = (\bar{\B}^{T} \bar{\R}^{-1} \bar{\B} + \Gam^{-1})^{-1}. 
\end{align}
Using the chain rule, the first derivative of $f(\boldsymbol{\phi})$  w.r.t. the $k$-th element $\phi_k$ can be expressed as
\begin{align}\label{f_der_phi}
\frac{\partial f(\boldsymbol{\phi})}{\partial \phi_k} = \text{tr}\Big\{  \Big(\frac{\partial f(\boldsymbol{\phi})}{\partial \X} \Big)^{T} \frac{\partial \X}{\partial \phi_k} \Big\}
\end{align}
where 
\begin{align}\label{f_der_X}
\frac{\partial f(\boldsymbol{\phi})}{\partial \X} = -\frac{\partial\text{tr}\{ \X\bar{\R}^{-1} \bar{\r}\bar{\r}^{T} \bar{\R}^{-1} \}}{\partial \X} = -\bar{\R}^{-1} \bar{\r}\bar{\r}^{T} \bar{\R}^{-1}
\end{align}
\begin{align}
\frac{\partial \X}{\partial \phi_k} = \frac{\partial \bar{\B}}{\partial\phi_k}\H \bar{\B}^{T} + \bar{\B}\frac{\partial \H}{\partial\phi_k}\bar{\B}^{T} + \bar{\B}\H \frac{\partial \bar{\B}^{T}}{\partial\phi_k}
\end{align}
with 
\begin{align}\label{H_der_phi}
\frac{\partial \H}{\partial\phi_k} = -\H\Big( \frac{\partial \bar{\B}^{T}}{\partial\phi_k} \bar{\R}^{-1}\bar{\B} + \bar{\B}^{T}\bar{\R}^{-1}\frac{\partial \bar{\B}}{\partial\phi_k} \Big)\H.
\end{align}
Substituting \eqref{f_der_X}-\eqref{H_der_phi} to \eqref{f_der_phi} and stacking the results in a vector form yields  \eqref{f_der_DOA}.

\end{document}